\documentclass[journal=jacsat,manuscript=article]{achemso}

\usepackage[version=3]{mhchem} 
\usepackage{siunitx}
\usepackage{xcolor} 
\usepackage[normalem]{ulem} 
\usepackage{graphicx}
\usepackage{tikz}
\usepackage{caption}
\usepackage{float}   
\usepackage{subfiles}
\usepackage{subcaption}

\newcommand{\equaltext}{These authors contributed equally to this work.}

\author{Juan Hurtado-Gallego}
\altaffiliation{\equaltext}
\affiliation[Université Catholique de Louvain (UCLouvain)]
{Institute of Condensed Matter and Nanosciences (IMCN/NAPS), Université Catholique de Louvain (UCLouvain), 1348 Louvain-la-Neuve, Belgium}

\author{Jérémie Pirard}
\altaffiliation{\equaltext}
\affiliation[Université Catholique de Louvain (UCLouvain)]
{Institute of Condensed Matter and Nanosciences (IMCN/NAPS), Université Catholique de Louvain (UCLouvain), 1348 Louvain-la-Neuve, Belgium}
 
\author{Abdalghani H. S. Daaoub}
\altaffiliation{\equaltext}
\affiliation[University of Warwick]
{Quantum Device Modelling Group, School of Engineering, University of Warwick, Coventry CV4 7AL, U.K}

\author{Sara Sangtarash}
\affiliation[University of Warwick]
{Quantum Device Modelling Group, School of Engineering, University of Warwick, Coventry CV4 7AL, U.K}

\author{Charlotte Kress}
\affiliation[University of Basel]
{Department of Chemistry, University of Basel, St. Johanns-Ring 19, Basel, 4056, Switzerland}

\author{Marcel Mayor}
\affiliation[University of Basel]
{Department of Chemistry, University of Basel, St. Johanns-Ring 19, Basel, 4056, Switzerland}
\alsoaffiliation[Karlsruhe Institute of Technology (KIT)]
{Institute for Nanotechnology, Karlsruhe Institute of Technology (KIT), P. O. Box 3640, Karlsruhe, 76021, Germany}
\alsoaffiliation[Sun Yat-Sen University]
{Lehn Institute of Functional Materials, School of Chemistry, Sun Yat-Sen University, St. Johanns-Ring 19, Guangzhou, 510274, P. R. China}

\author{Hatef Sadeghi}
\email{Hatef.Sadeghi@warwick.ac.uk} 
\affiliation[University of Warwick]
{Quantum Device Modelling Group, School of Engineering, University of Warwick, Coventry CV4 7AL, U.K}

\author{Pascal Gehring}
\email{Pascal.gehring@uclouvain.be} 
\affiliation[Université Catholique de Louvain (UCLouvain)]
{Institute of Condensed Matter and Nanosciences (IMCN/NAPS), Université Catholique de Louvain (UCLouvain), 1348 Louvain-la-Neuve, Belgium}
\alsoaffiliation[WEL]
{WEL Research Institute, Avenue Pasteur 6, 1300 Wavre, Belgium}

\title{Probing the dynamics and configurations of single molecule junctions via Seebeck coefficient spectroscopy}

\abbreviations{STM,$G$,$S$,DFT,$E_F$}

\begin{document}

\newpage
\begin{abstract}
  Single-molecule junctions exhibit dynamic structural configurations that strongly influence their electronic and thermoelectric properties. Here, we combine conductance ($G$) and Seebeck coefficient ($S$) measurements using the novel AC-based scanning tunnelling microscope break-junction technique to probe the real-time evolution of oligo(phenylene ethynylene) molecular junctions. We show that most junctions undergo configuration changes that lead to notable changes in $S$ while $G$ remains nearly constant. Density functional theory and quantum transport simulations link these observations to variations in contact geometry and charge transfer at the molecule–electrode interface. Our results demonstrate that simultaneous $G$ and $S$ measurements enable direct access to the dynamic reconfiguration of single-molecule junctions and offer design insights for thermoelectric molecular devices and new routes for increasing single-molecule junction stability.
\end{abstract}

\section{Introduction}

Break-junction techniques have evolved from simple transport measurements into powerful spectroscopic tools for probing electronic properties at the atomic scale \cite{Cuevas2017,AGRAIT200381,Gehring_2019,Schwarz_2014}. In molecular electronics, they are widely used to measure single-molecule conductance, but the results often exhibit broad statistical distributions due to the sensitivity of conductance ($G$) to molecular configuration, binding geometry, coupling strength, or strain. To study these effects in detail, large datasets are typically acquired and analyzed using unsupervised clustering methods \cite{Cabosart2019,Lemmer2016,Zotti2019}. However, a fundamental limitation remains: $G$ measurements probe only the transmission function at the Fermi level ($E_F$), making it difficult to distinguish between molecular conformations that yield similar conductance values \cite{Ornago2024}.

Thermoelectric measurements, particularly Seebeck coefficient ($S$) measurements, have emerged as a complementary approach to probe electronic transport in molecular junctions \cite{Rincon2016}. Unlike $G$, the Seebeck coefficient is sensitive to the slope of the transmission function at $E_F$, providing additional information about the electronic structure and energy dependence of transport channels. This approach has enabled the identification of whether charge transport through a single molecule is dominated by contributions from the highest occupied molecular orbital (HOMO) or the lowest unoccupied molecular orbital (LUMO) \cite{Cuevas2017}, as well as the role of chemical anchor groups \cite{Widawsky}, doping effects \cite{Baheti}, and the engineering of sharp transmission resonances that yield high thermoelectric performance \cite{Rincon2016_fullerene}. Recently, mechanical modulation experiments at ambient conditions demonstrated that simultaneous measurement of $G$ and $S$ is key to revealing subtle quantum interference effects, showing the destructive interference dip in the transmission function as a function of displacement \cite{Poel2024}. Moreover, fundamental molecular properties have been probed at cryogenic temperatures using thermocurrent spectroscopy, which has uncovered the spin ground state of radical molecules \cite{Eugenia2021}, revealed universal energy scales in the Kondo regime \cite{Bras2025}, and enabled the study of quantum phase transitions in single-molecule devices coupled to superconducting electrodes\cite{Volosheniuk2025}.

In this work, we leverage the simultaneous measurement of $G$ and $S$ as a function of displacement and time to directly capture molecular junction reconfigurations at the atomic level. By combining these two complementary transport properties with in-depth quantum transport and density functional theory (DFT) calculations, we gain new insight into the structural dynamics and electronic transport mechanisms of single-molecule junctions.

\section{Results and Discussion}

%

\begin{figure}[H]%
\centering
\includegraphics[width=1.0\textwidth]{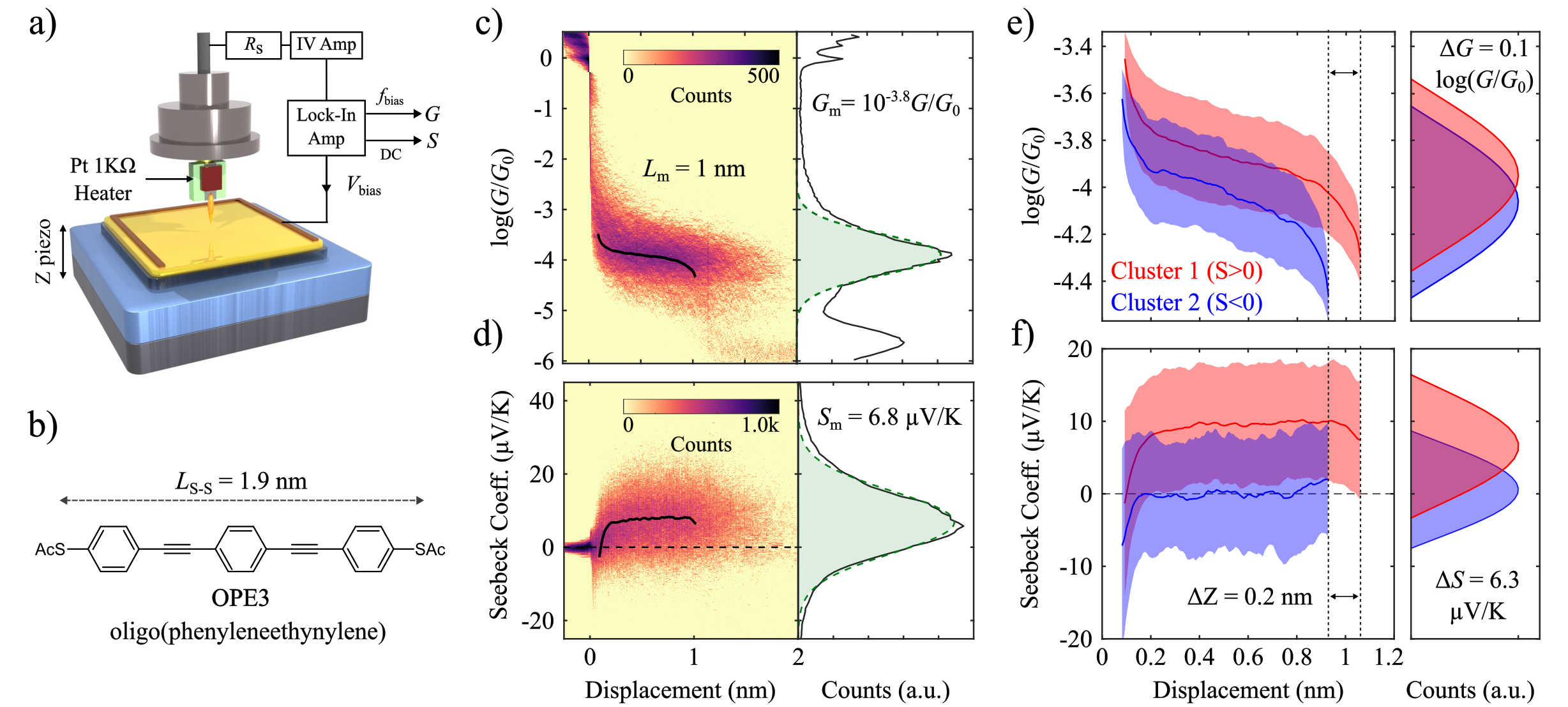}
\caption{ \textbf{a)} Schematic of the AC-STM setup. \textbf{b)} Structure of the acetyl masked rod-type model compound 1,4-bis(2'-(4''-mercaptophenyl)ethynyl)benzene (refered to as OPE3). Note that at the Au electrodes the acetyl masking groups are cleaved and Au-S bonds are formed. $L_\text{S-S}$ denotes the theoretical distance between the two sulfur atoms. 
\textbf{c)} Conductance (\textit{G}) measurement of OPE3. Left panel: 2D histogram of \textit{G} vs.\ displacement for all \textit{GZ} traces with a \textit{G} plateau. The black line on top of the 2D histogram represents the mean \textit{G} along the junction. \textit{$G_m$} and \textit{$L_m$} denote the mean molecular conductance and the mean junction length, respectively. Right panel: 1D histogram of \textit{G} for all \textit{GZ} traces with a \textit{G} plateau. The green dashed lines represents the Gaussian fit of the main peak of the 1D histogram. 
\textbf{d)} Same as in (c), but for the Seebeck coefficient \textit{S}. $S_m$ denotes the mean Seebeck coefficient extracted from the Gaussian fit. 
\textbf{e)} Left panel: \textit{G} vs.\ displacement for two selected clusters with positive (red) and negative (blue) \textit{S}. Solid lines represent the mean \textit{G} along the junction displacement, and the shaded regions indicate the standard deviation. Vertical dotted lines mark the mean length of each cluster; their difference in length is denoted by $\Delta Z$. Right panel: upper halves of the normalized Gaussian distributions. 
\textbf{f)} Same as in (d), but for \textit{S}.} 
\label{fig:STM_IZ_Traces}
\end{figure}

A home-made scanning tunnelling microscope (STM) was used to perform $G$ and $S$ measurements, using a freshly cut Au wire as a tip and 200nm thick Au(111) films on Mica as substrates. To form a single-molecule junction, the STM tip is repeatedly approached to and retracted from the sample surface. During retraction, a molecule can bridge the two electrodes, giving rise to a characteristic plateau in the conductance versus displacement trace, $G(z)$. This plateau serves as the primary signature of successful single-molecule junction formation. In our experiments we further measure $G$ simultaneously with  $S$, employing our recently developed AC-STM technique~\cite{Poel2024} (Fig.~\ref{fig:STM_IZ_Traces}a). An AC bias voltage ($V_{\mathrm{bias}}$) with frequency $f_{\mathrm{bias}} = \SI{3.123}{\kilo\hertz}$ and RMS amplitude of $25\text{mV}$ is applied to the sample, and the resulting AC current ($I_{\mathrm{SD}}$) is detected using standard lock-in techniques. Simultaneously, a temperature gradient ($\Delta T\approx30\text{K}$) is established between the tip and the sample by Joule heating a \SI{1}{\kilo\ohm} platinum resistor mounted on the tip, while the sample is maintained at room temperature. This temperature difference drives a DC thermoelectric current ($I_{\mathrm{th}}$), which is measured concurrently with $I_{\mathrm{SD}}$ (see SI). From these signals, $G$ and $S$ are extracted using $G = I_{\mathrm{SD}} / V_{\mathrm{bias}}$ and $S = I_{\mathrm{th}} / (G \Delta T)$, respectively (see SI for details). 

This method is applied to oligo(phenylene ethynylene) (OPE3) molecules (Fig.~\ref{fig:STM_IZ_Traces}b) functionalized with thioacetate (SAc) anchoring groups. In a typical experiment, we record thousands of individual displacement-dependent $G$ and $S$ traces. To analyze the data, we apply an unsupervised clustering technique \cite{Zotti2019,Cabosart2019,Lemmer2016} based on the \textit{k-means} algorithm (see SI for details), which allows us to distinguish \textit{GZ} traces that exhibit conductance plateaus (indicative of molecular junctions) from those showing only tunneling current (empty junctions). We then use these pre-selected subsets of experimental data to construct $G(z)$ and $S(z)$ histograms.

The resulting histograms (together with averaged curves in black) for the OPE3 molecule, based on approximately 2.100 individual traces, are shown in the left panels of Fig. \ref{fig:STM_IZ_Traces}c and Fig. \ref{fig:STM_IZ_Traces}d. The right panels display the corresponding 1D histograms--projections of the data onto the \textit{G} and \textit{S} axes--along with Gaussian distribution fits (dashed green lines). From the data, we determine a mean conductance of \( G_m = 10^{-3.8}G_0 \), a mean Seebeck coefficient of \( S_m = \SI{6.8}{\micro\volt\per\kelvin} \), and a mean molecular length of \( L_m = \SI{1}{\nano\meter} \) for the OPE3 molecule. In comparison, theoretical calculations performed with the Avogadro software predict a sulfur–sulfur distance of \( L_\text{S--S} = \SI{1.9}{\nano\meter} \). As previously noted~\cite{Arroyo2011}, experimental measurements of \( L_m \) generally underestimate the theoretical length, since the molecule is rarely found in a fully stretched configuration between the electrodes and mean length calculation does not favour this configuration. Our measured value of \( G_m \) is consistent with earlier reports for OPE3~\cite{Frisenda2016}.
In addition, the mean value of the Seebeck coefficient and its positive sign, which indicates a HOMO dominated transport through the junction\cite{Cuevas2017}, agree with values previously reported in literature using different techniques\cite{Miao2018,Gemma_2023}. Notably, we observe that $S(z)$ gradually increases from zero to its final mean value following junction formation. We attribute this evolution to dynamic coupling between the anchor groups and the electrodes during junction development\cite{Hurtado2024}.

It is important to highlight that the distribution of Seebeck coefficients (right panel of Fig.~\ref{fig:STM_IZ_Traces}d) exhibits a width of approximately 20~\si{\micro\volt\per\kelvin}. Such broad distributions are commonly observed in thermoelectric measurements and have been attributed to factors such as variations in contact geometry, intermolecular interactions, and torsional configurations of the molecular backbone~\cite{Malen2009}. Previous studies have been limited in their ability to probe the origin of this variability, as conventional measurement protocols typically yield only a single $S$ value per junction \cite{Widawsky,Reddy2007,Yee2011}. In contrast, our AC-based method enables the continuous tracking of $S$ throughout the formation, deformation, and rupture of the molecular junction. To gain further insight, we separate the individual \( G(z) \) traces by sorting them into two groups with average $S>0$ and $S<0$ (see SI for more details). The results are presented in Fig.~\ref{fig:STM_IZ_Traces}e and f. Two distinct clusters emerge in the Seebeck coefficient data (Fig.~\ref{fig:STM_IZ_Traces}f): one with a positive mean Seebeck coefficient (red) and another with a negative mean value (blue), with a difference of \( \Delta S = \SI{6.3}{\micro\volt\per\kelvin} \). The corresponding mean conductances for these two clusters are nearly identical (difference \( \Delta \log(G) = 0.1\, G/G_0 \), see Fig.~\ref{fig:STM_IZ_Traces}e), suggesting that the thermopower can vary significantly even among junctions with similar conductance characteristics. Lastly, we observe that the mean plateau lengths for the cluster with negative \( S \) values are approximately \( \Delta Z = \SI{0.2}{\nano\meter} \) shorter than those of the cluster with positive $S$.


\begin{figure}[H]%
\centering
\includegraphics[width=1\textwidth]{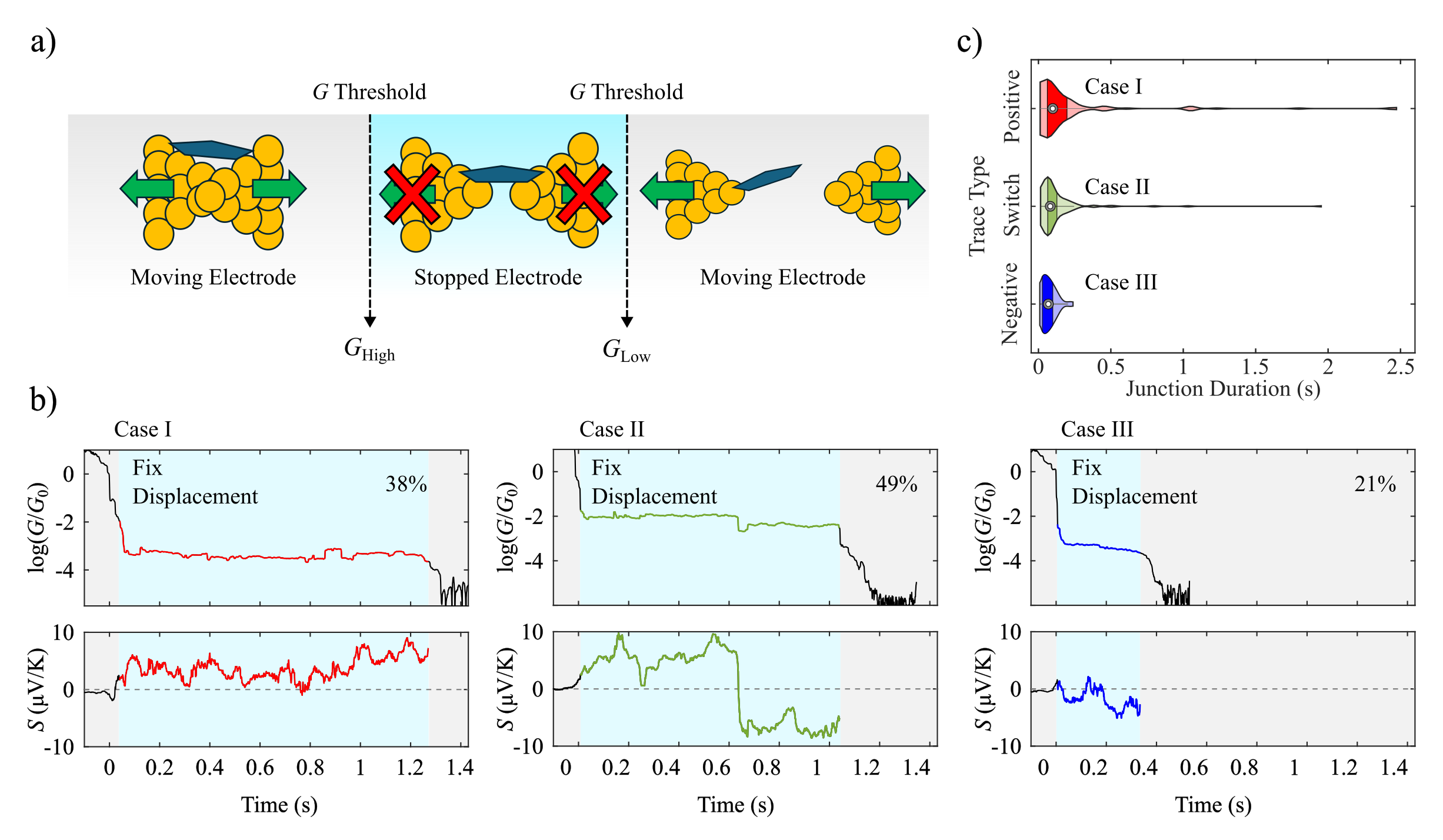}
\caption{\textbf{a)} Schematic of the self-breaking procedure. Three regions are distinguished: (i) Left, where the gold contact is still intact and the junction conductance \textit{G} exceeds $G_{High}$. In this regime, the molecule is located on the electrodes while they are being pulled apart. (ii) Middle, where \textit{G} lies between $G_{High}$ and $G_{Low}$. Here, the molecule bridges the tip and sample, and the electrode distance is held fixed. (iii) Right, where \textit{G} falls below $G_{Low}$, reaching the noise level of the system, and the electrodes are again moved apart. 
\textbf{b)} Examples of individual self-breaking traces, divided into three cases, with the percentage of traces in each group indicated at the top right. Blue shaded areas mark time intervals where the electrode distance is fixed, while grey shaded areas indicate when the electrodes are pulled apart. Left panel: Case I,  traces with positive \textit{S}. Middle panel: Case II, traces with switching between positive and negative \textit{S}. Right panel: Case III, traces with negative \textit{S}. 
\textbf{c)} Violin plots of junction durations for each cluster. The violin plots represent the kernel density estimate of the data. The colored box above each violin corresponds to the boxplot distribution, and the white points indicate the median junction duration for each case.} 
\label{fig:STM_self_freak_time}
\end{figure}

To investigate the various molecular configurations within the junction, we employed a self-breaking technique (see Fig.~\ref{fig:STM_self_freak_time}a) \cite{Huang2007,Tsutsui2009,Frisenda2015}. In this method, the junction is first opened until the conductance reaches an upper threshold \( G_{\mathrm{High}} > G_m \). At this point, the displacement is held constant, and both \textit{G} and \textit{S} are recorded as a function of time. If, during the self-breaking process, the conductance drops below a lower threshold \( G_{\mathrm{Low}} < G_m \), the motors are actuated to further open and fully break the junction. The junction is then closed again, and a new self-breaking cycle begins.

Fig.~\ref{fig:STM_self_freak_time}b shows three typical cases of time dependent $G$ and $S$ traces observed in such experiments (see Fig. S5-7 for more examples): case I (38\% of all traces, left panel) where $S$ stays positive over the whole junction evolution; case II (49\%, middle panel) where $S$ changes sign while the junction evolves; case III (21\%, right panel) where $S$ stays negative during self-breaking. The average conductances of each case $G_\mathrm{av}^I = 10^{-3.1}G_0$, $G_\mathrm{av}^{II} = 10^{-3.2}G_0$ and $G_\mathrm{av}^{III} = 10^{-3.9}G_0$ are very similar (see 2D histograms in Fig. S4). To assess the stability of different molecular configurations, we measured the junction duration time for each individual trace, defined as the time interval during which the electrode displacement is held constant. The resulting data are visualized in Fig.~\ref{fig:STM_self_freak_time}c using kernel density estimates, shown as violin plots for the three cases. The colored box on top of each violin represents the corresponding boxplot distribution, with white dots indicating the median junction duration for each cluster. Our analysis reveals that the most stable junctions correspond to traces with a constant positive Seebeck coefficient (\textit{S}, case I), whereas junctions exhibiting a negative \textit{S} (case III) tend to be less stable, exhibiting shorter lifetimes.


To investigate the electronic and thermoelectric properties of OPE3-based molecular junctions with varying contact geometries, we employed a combination of density functional theory (DFT) and quantum transport calculations \cite{Soler_2002,Sadeghi_2018}. This approach enables us to quantify the impact of structural variations at the molecule–electrode interface arising from forming various junctions in break-junction experiments on charge transport and thermoelectric performance.

\begin{figure}[H]%
\centering
\includegraphics[width=1\textwidth]{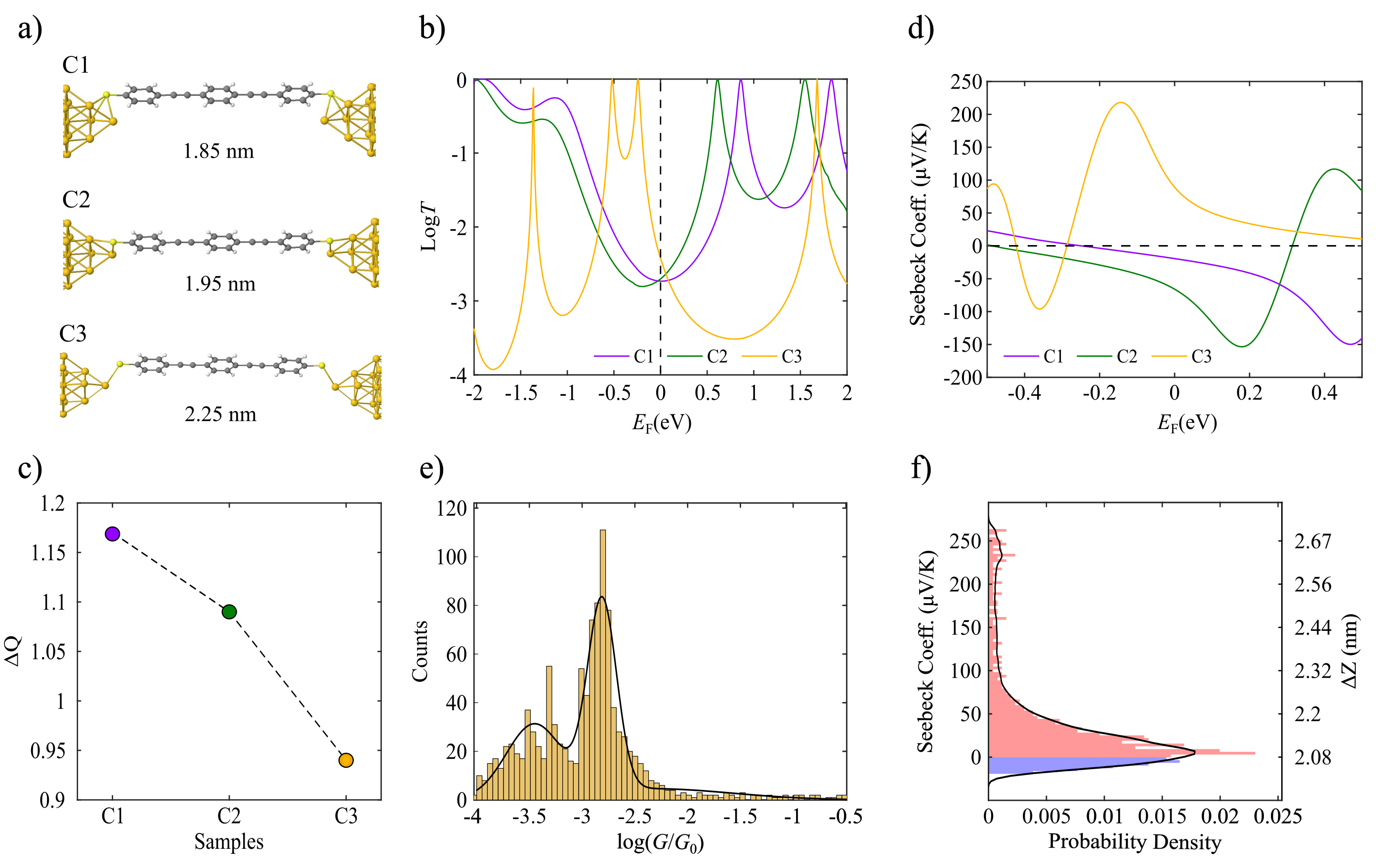}
\caption{Quantum transport through OPE3 molecular junctions with varying gold contact geometries. \textbf{a)} Optimized structures of oligo(phenylene ethynylene) (OPE3) molecule connected to gold electrodes through one (C3), two (C2), and three (C1) gold atoms at each contact point. \textbf{b)} Transmission coefficient (log $T$) as functions of energy for the three junction geometries, highlighting the effect of contact structure on electron transport. \textbf{c)} Calculated charge transfer from the gold electrodes to the OPE3 molecule for each contact configuration. \textbf{d)} Seebeck coefficients as functions of energy for each structure, corresponding to (b). \textbf{e)} Calculated room temperature conductance histograms over a range of molecule-electrode conformations over a Fermi energy range ($E_F$ = -0.1 to 0.7 eV), corresponding to Fig S12 (See SI for more information). \textbf{f)} Histogram of Seebeck coefficients sampled over the same Fermi energy range and structural conformations as in e.} 
\label{fig:Theory_Ex}
\end{figure}

We examined three realistic contact geometries, denoted C1, C2, and C3 (Fig.~\ref{fig:Theory_Ex}a), in which the OPE3 molecule is connected to gold electrodes via three, two, or one gold atom(s) at each side, respectively. These geometries were fully optimized prior to transport calculations. Material-specific mean-field Hamiltonians were extracted from DFT and integrated into the quantum transport code \textit{GOLLUM} \cite{Ferrer_2014,Sadeghi_2018} to calculate the energy-dependent transmission coefficient $T(E)$, from which electrical conductance and thermoelectric properties were obtained (see Computational Methods).

Fig.~\ref{fig:Theory_Ex}b shows the transmission function for three junctions shown in Fig.~\ref{fig:Theory_Ex}a. The transmission function is clearly affected by the changes on the molecule-electrode configuration. In particular the resonances are moved and the amplitude of transmission changes. In the C3 case, the HOMO resonance appears closer to $E_F$, enabling resonant transport. In contrast, in C1 and C2, the increased coupling shifts the LUMO closer to $E_F$, but also distorts its alignment, resulting in lower transmission near the DFT predicted Fermi level ($E=0$ eV). 

We attribute this to the changes in the charge transfer between the molecule and electrodes. As shown in Fig.~\ref{fig:Theory_Ex}c, the charge transfer ($\Delta Q$) from the electrodes to the molecule is highly sensitive to the contact geometry. The C1 configuration exhibits the greatest $\Delta Q$, followed by C2 and then C3. This trend is attributed to the increasing electrode–molecule contact area from C3 to C1, which enhances orbital overlap and electronic coupling. Consequently, in C1 and C2, the increased charge transfer shifts the molecular frontier orbitals (HOMO and LUMO) further below the electrode Fermi level, leading to a marked impact on transport behaviour.

This electronic reconfiguration significantly influences $S$, which depends on the slope of $T(E)$ at $E_F$ \cite{Sadeghi_2018}. Fig.~\ref{fig:Theory_Ex}d shows that C3 exhibits a positive and relatively large Seebeck coefficient due to a positive slope in $T(E)$, whereas both C1 and C2 exhibit negative values resulting from their negative slopes. Notably, $S$ in C3 exceeds that of C2 by approximately 70\%, reflecting the superior thermoelectric response in the weaker coupling regime.


To further assess the relationship between contact geometry, charge transfer, and transport behaviour, we considered junctions where OPE3 is connected via a single gold atom (C3-type), while systematically varying the tilt angle between the molecule and electrodes (See Figure S12 of the SI for details.) For each configuration, we computed $T(E)$ and the corresponding $G$ and $S$ as a function of the electrode's Fermi energy. Room-temperature $G$ histograms were generated (Fig.~\ref{fig:Theory_Ex}e) using the configurations in Figure S12 by sampling electrode Fermi levels over the range –0.1 to 0.7 eV. This range captures the relevant experimental bias window \cite{Sadeghi_2018,Neaton_2006} and accounts for the known underestimation of Fermi levels in standard DFT \cite{Zhao_2020,Hybertsen_1986,Ferri_2019}.

These histograms reveal a bimodal $G$ distribution \cite{Naghibi_2022}, associated with two structural regimes: (i) long, relaxed junctions with larger electrode–molecule separations, which exhibit higher $G$, and (ii) short, compressed junctions with lower $G$. Fig.~\ref{fig:Theory_Ex}f further correlates these regimes with thermopower: ~70\% of the sampled configurations (longer junctions) show positive $S$ values, whereas ~30\% (shorter junctions) exhibit negative ones. The average length difference between these two regimes is approximately 0.25 nm. These are in good agreement with our measured $G$ and $S$ histograms. This analysis highlights a consistent mechanism: as the junction becomes shorter and the tilt angle decreases, orbital overlap and charge transfer increase, leading to a downshift in molecular orbital energies and a change of the sign of $S$ due to changes in the slope of $T(E)$ close to $E_F$. These results highlight the added value of simultaneous \textit{G} and \textit{S} measurements. While conductance histograms and self-breaking experiments reveal minimal statistically significant variations in \( G \) between traces, the corresponding values of \( S \) can vary dramatically. The self-breaking analysis shows that the configuration yielding a positive \( S \)—commonly reported in the literature (case I)—corresponds to the most stable junction. However, the most statistically probable scenario in our measurements involves changes in the sign of \( S \) during the junction's lifetime (case II). This indicates that, under ambient conditions, most single-molecule junctions are dynamic, with the molecule exploring different configurations on a timescale of tens of milliseconds.

In summary, our study revealed that single-molecule junctions are inherently dynamic even at a fixed electrode separation, with the molecule adopting multiple configurations during measurement. Using the AC-STM-BJ technique, we simultaneously measured conductance and Seebeck coefficient without perturbing the junction. Density Functional Theory calculations support the presence of distinct molecular conformations consistent with experimental data. Additionally, subtle changes in contact geometry—such as bonding configuration and tilt angle—significantly influence charge transfer at the electrode–molecule interface, tuning orbital alignment, transmission spectra, conductance, and thermopower. These findings provide a mechanistic framework linking theory and experiment and offer insights for the rational design of stable and efficient thermoelectric molecular devices.

\section{Methods}
\textbf{ Sample preparation:}
To this end, molecules are deposited on top of freshly annealed Au(111) samples, by immersing the sample into a $1\text{mM}$ DCM solution, and freshly cut 0.2 mm diameter, 99.9\% pure Au wire is used as tip. 

\textbf{\textit{K-means} classification:}
In order to consider only the properties of the molecular junction, the \textit{G} and \textit{S} values included in the \textit{k-means} function are the ones placed in the \textit{G} range between \textit{$G_m$} $ \pm $ $ G_{std} $, where \textit{$G_m$} and $ G_{std} $ are the mean conductance and standard deviation of the respective Gaussian fits to the 1D \textit{G} histograms of each cluster.  

\textbf{ Sub-cluster classification:} To separate positive and negative \textit{S} traces, we first calculate the mean \textit{S} value for each individual trace. Traces with a mean \textit{S} greater than \(\SI{2}{\micro\volt\per\kelvin} \) are classified into the \( S > 0 \) cluster, while those with a mean \textit{S} less than \(\SI{2}{\micro\volt\per\kelvin} \) are assigned to the \( S < 0 \) cluster.

\textbf{ Computational:}The optimized geometry, ground state Hamiltonian and overlap matrix elements of each structure were self-consistently obtained using the SIESTA[1] implementation of density functional theory (DFT). SIESTA employs norm-conserving pseudo-potentials to account for the core electrons and linear combinations of atomic orbitals to construct the valence states. The local density approximation (GGA) of the exchange and correlation functional is used with PBE parameterization, a double-$\zeta$ polarized (DZP) basis set, a real-space grid defined with an equivalent energy cut-off of 250 Ry. The geometry optimization for each structure is performed to the forces smaller than 10 meV/Å. The mean-field Hamiltonian obtained from the converged DFT calculation was combined with the GOLLUM [2,3] implementation of the non-equilibrium Green’s function method to calculate the phase-coherent, elastic scattering properties of the each system consist of left gold (source) and right gold (drain) leads and the scattering region. The transmission coefficient T(E) for electrons of energy E (passing from the source to the drain) is calculated via the relation: $T(E) = \mathrm{Tr} \left[ \Gamma_R(E) G^R(E) \Gamma_L(E) G^{R\dagger}(E) \right]$. In this expression, $\Gamma_{L,R}(E) = i \left( \Sigma_{L,R}(E) - \Sigma_{L,R}^\dagger(E) \right)$ describe the level broadening due to the coupling between left (L) and right (R) electrodes and the central scattering region, $\Sigma_{L,R}(E)$  are the retarded self-energies associated with this coupling and $
G^R(E) = \left( E S - H - \Sigma_L(E) - \Sigma_R(E) \right)^{-1}
$ is the retarded Green’s function.

\textbf{Thermoelectric properties: }
The electrical conductance $G = G_0 L_0$ and the Seebeck coefficient $
S = -\frac{L_1}{e T L_0}$ are calculated from the electron transmission coefficient $T(E)$ where the momentums $L_n = \int_{-\infty}^{+\infty} dE \, (E - E_F)^n \, T(E) \left( -\frac{\partial f_{FD}}{\partial E} \right)$ and $f_{FD}$ is the Fermi-Dirac probability distribution function $
f_{FD} = \left( e^{\frac{E - E_F}{k_B T}} + 1 \right)^{-1}$, T is the temperature, $E_F$ is the Fermi energy, $G_0 = \frac{2 e^{2}}{h}$ is the conductance quantum, e is electron charge and h is the Planck’s constant. The calculated histograms are formed using the method explained in \cite{Daaoub_2022}. To calculate the distribution of charges on each molecule and their redistribution according to molecular conformations, we calculated the Mulliken charge on each atom for each conformation using DFT. The difference in charge is calculated as $
\Delta Q = | Q_n - Q_0 |
$ where $Q_n$ is the Mulliken charge on a given atom or group of atoms for the configuration n (between the gold electrodes) and $Q_0$ is the Mulliken charge at the same atom(s) in the ground state configuration of the isolated molecule (without the electrodes).
\begin{acknowledgement}

P.G and J.P acknowledge the financial support from the EU (ERC-StG-10104144-MOUNTAIN), from the F.R.S.-FNRS of Belgium (FNRS-CQ-1.C044.21-SMARD, FNRS-CDR-J.006823. F1-SiMolHeat).
J.H-G. is grateful to the UCLouvain for the award of an FSR Incoming Post-doc fellowship.
H.S. acknowledges UKRI for Future Leaders Fellowships MR/S015329/2 and MR/X015181/1. S.S. acknowledges EPSRC New Investigator Grant APP17327.
C.K. and M.M. acknowledge generous support from the Swiss National Science Foundation (SNF Grant no. 200020-207744). M.M. acknowledges support from the 111 project (Grant No. 90002-18011002).

\end{acknowledgement}

\begin{suppinfo}

Figures S1-S2 shows experimental set-up and calibration. Figure S3 shows clustering technique. Figures S4-S7 self-breaking technique results. Figures S8-S12 and Table 1 show computational calculations.

\end{suppinfo}

\section{Author Contribution} The manuscript was written through contributions of all authors. All authors have given approval to the final version of the manuscript. \textsuperscript{\#}J.H-G, J.P and A.D contribute equally to this work.

\bibliography{bib_OPE3}

@article{Lemmer2016,
   author = {Mario Lemmer and Michael S. Inkpen and Katja Kornysheva and Nicholas J. Long and Tim Albrecht},
   doi = {10.1038/ncomms12922},
   issn = {20411723},
   journal = {Nature Communications},
   month = {10},
   publisher = {Nature Publishing Group},
   title = {Unsupervised vector-based classification of single-molecule charge transport data},
   volume = {7},
   year = {2016},
}

@article{Cabosart2019,
   author = {Damien Cabosart and Maria El Abbassi and Davide Stefani and Riccardo Frisenda and Michel Calame and Herre S.J. Van der Zant and Mickael L. Perrin},
   doi = {10.1063/1.5089198},
   issn = {00036951},
   issue = {14},
   journal = {Applied Physics Letters},
   month = {4},
   publisher = {American Institute of Physics Inc.},
   title = {A reference-free clustering method for the analysis of molecular break-junction measurements},
   volume = {114},
   year = {2019},
}

@article{Widawsky,
author = {Widawsky, Jonathan R. and Darancet, Pierre and Neaton, Jeffrey B. and Venkataraman, Latha},
title = {Simultaneous Determination of Conductance and Thermopower of Single Molecule Junctions},
journal = {Nano Letters},
volume = {12},
number = {1},
pages = {354-358},
year = {2012},
doi = {10.1021/nl203634m},
    note ={PMID: 22128800},

URL = { 
    
        https://doi.org/10.1021/nl203634m
    
    

},
eprint = { 
    
        https://doi.org/10.1021/nl203634m
    
    

}

}

@article{Baheti,
author = {Baheti, Kanhayalal and Malen, Jonathan A. and Doak, Peter and Reddy, Pramod and Jang, Sung-Yeon and Tilley, T. Don and Majumdar, Arun and Segalman, Rachel A.},
title = {Probing the Chemistry of Molecular Heterojunctions Using Thermoelectricity},
journal = {Nano Letters},
volume = {8},
number = {2},
pages = {715-719},
year = {2008},
doi = {10.1021/nl072738l},
    note ={PMID: 18269258},

URL = { 
    
        https://doi.org/10.1021/nl072738l
    
    

},
eprint = { 
    
        https://doi.org/10.1021/nl072738l
    
    

}

}

@article{Zotti2019,
   author = {Linda A. Zotti and Beatrice Bednarz and Juan Hurtado-Gallego and Damien Cabosart and Gabino Rubio-Bollinger and Nicolas Agra\"it and Herre S.J. van der Zant},
   doi = {10.3390/biom9100580},
   issn = {2218273X},
   issue = {10},
   journal = {Biomolecules},
   month = {10},
   pmid = {31591358},
   publisher = {MDPI AG},
   title = {Can one define the conductance of amino acids?},
   volume = {9},
   year = {2019},
}

@article{AGRAIT200381,
title = {Quantum properties of atomic-sized conductors},
journal = {Physics Reports},
volume = {377},
number = {2},
pages = {81-279},
year = {2003},
issn = {0370-1573},
doi = {https://doi.org/10.1016/S0370-1573(02)00633-6},
url = {https://www.sciencedirect.com/science/article/pii/S0370157302006336},
author = {Nicolás Agra\"it and Alfredo Levy Yeyati and Jan M. {van Ruitenbeek}}
}

@article{
Reddy2007,
author = {Pramod Reddy  and Sung-Yeon Jang  and Rachel A. Segalman  and Arun Majumdar },
title = {Thermoelectricity in Molecular Junctions},
journal = {Science},
volume = {315},
number = {5818},
pages = {1568-1571},
year = {2007},
doi = {10.1126/science.1137149},
URL = {https://www.science.org/doi/abs/10.1126/science.1137149},
eprint = {https://www.science.org/doi/pdf/10.1126/science.1137149}}

@article{Yee2011,
author = {Yee, Shannon K. and Malen, Jonathan A. and Majumdar, Arun and Segalman, Rachel A.},
title = {Thermoelectricity in Fullerene–Metal Heterojunctions},
journal = {Nano Letters},
volume = {11},
number = {10},
pages = {4089-4094},
year = {2011},
doi = {10.1021/nl2014839},
    note ={PMID: 21882860},

URL = { 
    
        https://doi.org/10.1021/nl2014839
    
    

},
eprint = { 
    
        https://doi.org/10.1021/nl2014839
    
    

}

}

@article{Volosheniuk2025,
   author = {Volosheniuk, Serhii and Bouwmeester, Damian and Vogel, David and Wegeberg, Christina and Hsu, Chunwei and Mayor, Marcel and van der Zant, Herre S. J. and Gehring, Pascal},
   title = {Enhancing thermoelectric output in a molecular heat engine utilizing Yu-Shiba-Rusinov bound states},
   journal = {Nature Communications},
   volume = {16},
   number = {1},
   pages = {3279},
   ISSN = {2041-1723},
   DOI = {10.1038/s41467-025-58645-1},
   url = {https://doi.org/10.1038/s41467-025-58645-1},
   year = {2025},
   type = {Journal Article}
}

@article{Bras2025,
author = {Bras, Tristan and Hsu, Chunwei and Baum, Thomas Y. and Vogel, David and Mayor, Marcel and van der Zant, Herre S. J.},
title = {Mechanically Stable Kondo Resonance in an Organic Radical Molecular Junction},
journal = {The Journal of Physical Chemistry C},
volume = {129},
number = {6},
pages = {3152-3157},
year = {2025},
doi = {10.1021/acs.jpcc.4c05860},

URL = { 
    
        https://doi.org/10.1021/acs.jpcc.4c05860
    
    

},
eprint = { 
    
        https://doi.org/10.1021/acs.jpcc.4c05860
    
    

}

}

@article{Eugenia2021,
author = {Pyurbeeva, Eugenia and Hsu, Chunwei and Vogel, David and Wegeberg, Christina and Mayor, Marcel and van der Zant, Herre and Mol, Jan A. and Gehring, Pascal},
title = {Controlling the Entropy of a Single-Molecule Junction},
journal = {Nano Letters},
volume = {21},
number = {22},
pages = {9715-9719},
year = {2021},
doi = {10.1021/acs.nanolett.1c03591},
    note ={PMID: 34766782},

URL = { 
    
        https://doi.org/10.1021/acs.nanolett.1c03591
    
    

},
eprint = { 
    
        https://doi.org/10.1021/acs.nanolett.1c03591
    
    

}

}

@article{Rincon2016_fullerene,
   author = {Rincón-García, Laura and Ismael, Ali K. and Evangeli, Charalambos and Grace, Iain and Rubio-Bollinger, Gabino and Porfyrakis, Kyriakos and Agra\"it, Nicolás and Lambert, Colin J.},
   title = {Molecular design and control of fullerene-based bi-thermoelectric materials},
   journal = {Nature Materials},
   volume = {15},
   number = {3},
   pages = {289-293},
   ISSN = {1476-4660},
   DOI = {10.1038/nmat4487},
   url = {https://doi.org/10.1038/nmat4487},
   year = {2016},
   type = {Journal Article}
}

@article{Ornago2024,
author = {Ornago, Luca and Zwick, Patrick and van der Poel, Sebastiaan and Brandl, Thomas and El Abbassi, Maria and Perrin, Mickael L. and Dulić, Diana and van der Zant, Herre S. J. and Mayor, Marcel},
title = {Influence of Peripheral Alkyl Groups on Junction Configurations in Single-Molecule Electronics},
journal = {The Journal of Physical Chemistry C},
volume = {128},
number = {3},
pages = {1413-1422},
year = {2024},
doi = {10.1021/acs.jpcc.3c06970},

URL = { 
    
        https://doi.org/10.1021/acs.jpcc.3c06970
    
    

},
eprint = { 
    
        https://doi.org/10.1021/acs.jpcc.3c06970
    
    

}

}

@article{Rincon2016,
   author = {Laura Rincón-García and Charalambos Evangeli and Gabino Rubio-Bollinger and Nicolás Agra\"it},
   doi = {10.1039/c6cs00141f},
   issn = {14604744},
   issue = {15},
   journal = {Chemical Society Reviews},
   month = {8},
   pages = {4285-4306},
   publisher = {Royal Society of Chemistry},
   title = {Thermopower measurements in molecular junctions},
   volume = {45},
   year = {2016},
}

@article{Poel2024,
   author = {Sebastiaan van der Poel and Juan Hurtado-Gallego and Matthias Blaschke and Rubén López-Nebreda and Almudena Gallego and Marcel Mayor and Fabian Pauly and Herre S.J. van der Zant and Nicolás Agra\"it},
   doi = {10.1038/s41467-024-53825-x},
   issn = {20411723},
   issue = {1},
   journal = {Nature Communications },
   month = {12},
   publisher = {Nature Research},
   title = {Mechanoelectric sensitivity reveals destructive quantum interference in single-molecule junctions},
   volume = {15},
   year = {2024},
}

@article{Hurtado2024,
   author = {Juan Hurtado-Gallego and Sebastiaan van der Poel and Matthias Blaschke and Almudena Gallego and Chunwei Hsu and Rubén López-Nebreda and Marcel Mayor and Fabian Pauly and Nicolás Agra\"it and Herre S.J. van der Zant},
   doi = {10.1039/d4nr00704b},
   issn = {20403372},
   issue = {22},
   journal = {Nanoscale},
   month = {4},
   pages = {10751-10759},
   pmid = {38747099},
   publisher = {Royal Society of Chemistry},
   title = {Benchmarking break-junction techniques: electric and thermoelectric characterization of naphthalenophanes},
   volume = {16},
   year = {2024},
}

@article{Frisenda2016,
   author = {Riccardo Frisenda and Vera A.E.C. Janssen and Ferdinand C. Grozema and Herre S.J. Van Der Zant and Nicolas Renaud},
   doi = {10.1038/nchem.2588},
   issn = {17554349},
   issue = {12},
   journal = {Nature Chemistry},
   month = {12},
   pages = {1099-1104},
   publisher = {Nature Publishing Group},
   title = {Mechanically controlled quantum interference in individual {$\pi$}-stacked dimers},
   volume = {8},
   year = {2016},
}

@article{Miao2018,
   author = {Ruijiao Miao and Hailiang Xu and Maxim Skripnik and Longji Cui and Kun Wang and Kim G.L. Pedersen and Martin Leijnse and Fabian Pauly and Kenneth Wärnmark and Edgar Meyhofer and Pramod Reddy and Heiner Linke},
   doi = {10.1021/acs.nanolett.8b02207},
   issn = {15306992},
   issue = {9},
   journal = {Nano Letters},
   month = {9},
   pages = {5666-5672},
   pmid = {30084643},
   publisher = {American Chemical Society},
   title = {Influence of Quantum Interference on the Thermoelectric Properties of Molecular Junctions},
   volume = {18},
   year = {2018},
}

@article{Malen2009,
   author = {Jonathan A. Malen and Peter Doak and Kanhayalal Baheti and T. Don Tilley and Arun Majumdar and Rachel A. Segalman},
   doi = {10.1021/nl9013875},
   issn = {15306984},
   issue = {10},
   journal = {Nano Letters},
   month = {10},
   pages = {3406-3412},
   title = {The nature of transport variations in molecular heterojunction electronics},
   volume = {9},
   year = {2009},
}

@article{Huang2007,
   author = {Zhifeng Huang and Fang Chen and Peter A. Bennett and Nongjian Tao},
   doi = {10.1021/ja074456t},
   issn = {00027863},
   issue = {43},
   journal = {Journal of the American Chemical Society},
   month = {10},
   pages = {13225-13231},
   pmid = {17915870},
   title = {Single molecule junctions formed via Au-thiol contact: Stability and breakdown mechanism},
   volume = {129},
   year = {2007},
}

@article{Tsutsui2009,
   author = {Makusu Tsutsui and Masateru Taniguchi and Tomoji Kawai},
   doi = {10.1021/ja902871d},
   issn = {00027863},
   issue = {30},
   journal = {Journal of the American Chemical Society},
   month = {8},
   pages = {10552-10556},
   title = {Quantitative evaluation of metal-molecule contact stability at the single-molecule level},
   volume = {131},
   year = {2009},
}

@article{Frisenda2015,
   author = {Riccardo Frisenda and Simge Tarkuç and Elena Galán and Mickael L. Perrin and Rienk Eelkema and Ferdinand C. Grozema and Herre S.J. van der Zant},
   doi = {10.3762/bjnano.6.159},
   issn = {21904286},
   issue = {1},
   journal = {Beilstein Journal of Nanotechnology},
   pages = {1558-1567},
   publisher = {Beilstein-Institut Zur Forderung der Chemischen Wissenschaften},
   title = {Electrical properties and mechanical stability of anchoring groups for single-molecule electronics},
   volume = {6},
   year = {2015},
}

@article{Arroyo2011,
author = {Arroyo, Carlos R. and Leary, Edmund and Castellanos-G\'omez, Andr{\'e}s and Rubio-Bollinger, Gabino and González, M. Teresa and Agra\"it, Nicolás},
title = {Influence of Binding Groups on Molecular Junction Formation},
journal = {Journal of the American Chemical Society},
volume = {133},
number = {36},
pages = {14313-14319},
year = {2011},
doi = {10.1021/ja201861k},
    note ={PMID: 21806051},

URL = { 
    
        https://doi.org/10.1021/ja201861k
    
    

},
eprint = { 
    
        https://doi.org/10.1021/ja201861k
    
    

}

}

@book{Cuevas2017,
   author = {Juan Carlos Cuevas and Elke Scheer},
   edition = {2nd},
   pages = {848},
   publisher = {World Scientific Series in Nanoscience and Nanotechnology},
   title = {Molecular Electronics:
An Introduction to Theory and Experiment,
2nd Edition},
   year = {2017},
}

@article{Soler_2002,
   title={The SIESTA method for ab initio order-N materials simulation},
   volume={14},
   ISSN={1361-648X},
   url={http://dx.doi.org/10.1088/0953-8984/14/11/302},
   DOI={10.1088/0953-8984/14/11/302},
   number={11},
   journal={Journal of Physics: Condensed Matter},
   publisher={IOP Publishing},
   author={Soler, José M and Artacho, Emilio and Gale, Julian D and García, Alberto and Junquera, Javier and Ordejón, Pablo and Sánchez-Portal, Daniel},
   year={2002},
   month=mar, pages={2745–2779} }

@article{Ferrer_2014,
   title={GOLLUM: a next-generation simulation tool for electron, thermal and spin transport},
   volume={16},
   ISSN={1367-2630},
   url={http://dx.doi.org/10.1088/1367-2630/16/9/093029},
   DOI={10.1088/1367-2630/16/9/093029},
   number={9},
   journal={New Journal of Physics},
   publisher={IOP Publishing},
   author={Ferrer, J and Lambert, C J and García-Suárez, V M and Manrique, D Zs and Visontai, D and Oroszlany, L and Rodríguez-Ferradás, R and Grace, I and Bailey, S W D and Gillemot, K and Sadeghi, Hatef and Algharagholy, L A},
   year={2014},
   month=sep, pages={093029} }

@article{Sadeghi_2018,
doi = {10.1088/1361-6528/aace21},
url = {https://dx.doi.org/10.1088/1361-6528/aace21},
year = {2018},
month = {jul},
publisher = {IOP Publishing},
volume = {29},
number = {37},
pages = {373001},
author = {Sadeghi, Hatef},
title = {Theory of electron, phonon and spin transport in nanoscale quantum devices},
journal = {Nanotechnology}
}

@article{Neaton_2006,
   title={Renormalization of Molecular Electronic Levels at Metal-Molecule Interfaces},
   volume={97},
   ISSN={1079-7114},
   url={http://dx.doi.org/10.1103/PhysRevLett.97.216405},
   DOI={10.1103/physrevlett.97.216405},
   number={21},
   journal={Physical Review Letters},
   publisher={American Physical Society (APS)},
   author={Neaton, J. B. and Hybertsen, Mark S. and Louie, Steven G.},
   year={2006},
pages = {216405},
   month=nov }

@article{Zhao_2020,
author = {Zhao, Zhi-Hao and Wang, Lin and Li, Shi and Zhang, Wei-Dong and He, Gang and Wang, Dong and Hou, Shi-Min and Wan, Li-Jun},
title = {Single-Molecule Conductance through an Isoelectronic B–N Substituted Phenanthrene Junction},
journal = {Journal of the American Chemical Society},
volume = {142},
number = {18},
pages = {8068-8073},
year = {2020},
doi = {10.1021/jacs.0c00879},
    note ={PMID: 32321243},
URL = { 
        https://doi.org/10.1021/jacs.0c00879
},
eprint = {  
        https://doi.org/10.1021/jacs.0c00879
}
}

@article{Hybertsen_1986,
  title = {Electron correlation in semiconductors and insulators: Band gaps and quasiparticle energies},
  author = {Hybertsen, Mark S. and Louie, Steven G.},
  journal = {Phys. Rev. B},
  volume = {34},
  issue = {8},
  pages = {5390--5413},
  numpages = {0},
  year = {1986},
  month = {Oct},
  publisher = {American Physical Society},
  doi = {10.1103/PhysRevB.34.5390},
  url = {https://link.aps.org/doi/10.1103/PhysRevB.34.5390}
}

@article{Ferri_2019,
author = {Ferri, Nicolò and Algethami, Norah and Vezzoli, Andrea and Sangtarash, Sara and McLaughlin, Maeve and Sadeghi, Hatef and Lambert, Colin J. and Nichols, Richard J. and Higgins, Simon J.},
title = {Hemilabile Ligands as Mechanosensitive Electrode Contacts for Molecular Electronics},
journal = {Angewandte Chemie International Edition},
volume = {58},
number = {46},
pages = {16583-16589},
doi = {https://doi.org/10.1002/anie.201906400},
url = {https://onlinelibrary.wiley.com/doi/abs/10.1002/anie.201906400},
eprint = {https://onlinelibrary.wiley.com/doi/pdf/10.1002/anie.201906400},
year = {2019}
}

@article{Naghibi_2022,
author = {Naghibi, Saman and Sangtarash, Sara and Kumar, Varshini J. and Wu, Jian-Zhong and Judd, Martyna M. and Qiao, Xiaohang and Gorenskaia, Elena and Higgins, Simon J. and Cox, Nicholas and Nichols, Richard J. and Sadeghi, Hatef and Low, Paul J. and Vezzoli, Andrea},
title = {Redox-Addressable Single-Molecule Junctions Incorporating a Persistent Organic Radical},
journal = {Angewandte Chemie International Edition},
volume = {61},
number = {23},
pages = {e202116985},
doi = {https://doi.org/10.1002/anie.202116985},
url = {https://onlinelibrary.wiley.com/doi/abs/10.1002/anie.202116985},
eprint = {https://onlinelibrary.wiley.com/doi/pdf/10.1002/anie.202116985},
year = {2022}
}

@article{Daaoub_2022,
author = {Daaoub, Abdalghani and Ornago, Luca and Vogel, David and Bastante, Pablo and Sangtarash, Sara and Parmeggiani, Matteo and Kamer, Jerry and Agra\"it, Nicolás and Mayor, Marcel and van der Zant, Herre and Sadeghi, Hatef},
title = {Engineering Transport Orbitals in Single-Molecule Junctions},
journal = {The Journal of Physical Chemistry Letters},
volume = {13},
number = {39},
pages = {9156-9164},
year = {2022},
doi = {10.1021/acs.jpclett.2c01851},
    note ={PMID: 36166407},
URL = { 
        https://doi.org/10.1021/acs.jpclett.2c01851
},
eprint = { 
        https://doi.org/10.1021/acs.jpclett.2c01851
}
}

@article{Gemma_2023,
   author = {Gemma, Andrea and Tabatabaei, Fatemeh and Drechsler, Ute and Zulji, Anel and Dekkiche, Hervé and Mosso, Nico and Niehaus, Thomas and Bryce, Martin R. and Merabia, Samy and Gotsmann, Bernd},
   title = {Full thermoelectric characterization of a single molecule},
   journal = {Nature Communications},
   volume = {14},
   number = {1},
   pages = {3868},
   ISSN = {2041-1723},
   DOI = {10.1038/s41467-023-39368-7},
   url = {https://doi.org/10.1038/s41467-023-39368-7},
   year = {2023},
   type = {Journal Article}
}

@article{Gehring_2019,
   author = {Gehring, Pascal and Thijssen, Jos M. and van der Zant, Herre S. J.},
   title = {Single-molecule quantum-transport phenomena in break junctions},
   journal = {Nature Reviews Physics},
   volume = {1},
   number = {6},
   pages = {381-396},
   ISSN = {2522-5820},
   DOI = {10.1038/s42254-019-0055-1},
   url = {https://doi.org/10.1038/s42254-019-0055-1},
   year = {2019},
   type = {Journal Article}
}

@article{Schwarz_2014,
doi = {10.1088/0953-8984/26/47/474201},
url = {https://dx.doi.org/10.1088/0953-8984/26/47/474201},
year = {2014},
month = {oct},
publisher = {IOP Publishing},
volume = {26},
number = {47},
pages = {474201},
author = {Schwarz, Florian and L\"ortscher, Emanuel},
title = {Break-junctions for investigating transport at the molecular scale},
journal = {Journal of Physics: Condensed Matter}
}
\end{document}